\documentclass[11pt]{article}

\usepackage[english]{babel}								
\usepackage[utf8]{inputenc}									 
\usepackage{amsmath,amsfonts,amssymb,amsthm,cancel,siunitx,
	calculator,calc,mathtools,empheq,latexsym}

\usepackage[hidelinks]{hyperref}
\hypersetup{colorlinks=true,linkcolor=blue, citecolor=blue, urlcolor=blue}
\usepackage{titlesec}
\usepackage{lmodern}
\titleformat*{\section}{\large\bfseries}
\allowdisplaybreaks
\newcommand{\Ffourthree}[8]{%
	\,{}_4F_3\biggl(%
	\genfrac{}{}{0pt}{}{#1\,,\:#2\,,\:#3\,,\:#4}{#5\,,\:#6\,,\:#7}
	\bigg\vert#8\biggr)}

\title{\large\bf
	{Celestial amplitudes dual to the $\mathbf{O\left(N\right)}$ nonlinear sigma model}}

\author{Valeriia Stolbova }	

\begin{document}
\numberwithin{equation}{section}
\thispagestyle{empty}
	\date{}
	\maketitle
	\vspace{-1cm}
	\begin{center}
		{\normalsize
			Steklov Mathematical Institute, Fontanka 27, St. Petersburg, 191023, Russia \\
			\vspace{5mm}
			E-mail: stolbova@pdmi.ras.ru}
	\end{center}
	\bigskip
	\noindent
	{\small \centering{{\bf Abstract}}\\
        We compute celestial amplitudes corresponding to the exact S-matrix of the $2d$ $O(N)$-symmetric nonlinear sigma model. Celestial amplitudes for two-dimensional integrable S-matrices simplify to Fourier transforms. Due to the connection between Fourier and Mellin transforms, celestial amplitudes for $O(N)$ model are Mejer G-functions. We also prove crossing symmetry for obtained results and review simplifications for $O(3)$ symmetry case.}	
    
	
%\baselineskip=\normalbaselineskip
%\tableofcontents
\small\tableofcontents
\normalsize

\section{Introduction}
Celestial conformal field theory, or CCFT, is one of the modern approaches to flat space holography, which is still a developing area \cite{strom_lect, past_lect}. This approach is based on the fact that the $\left(d+2\right)$-dimensional Lorentz group is isomorphic to the $d$-dimensional Euclidean conformal group \cite{conf_sym}. It follows that there is a certain basis of wavefunctions called conformal basis where scattering amplitudes in $\mathbb{R}^{1,d+1}$ admit interpretations as Euclidean $d$-dimensional conformal correlators \cite{conf_bas}. The discovery of this basis is motivated in particular by the search of two-dimensional symmetries in four-dimensional scattering amplitudes \cite{hol, boer_solod}.

A class of quantum field theory models with non-trivial symmetries is that of two-dimensional integrable models \cite{bomb}. Due to the presence of high-order conserved charges, integrable S-matrices in $2d$ possess the factorization property \cite{parke, iag}, together with the physical unitarity and crossing symmetry properties \cite{muss}. These symmerties make it possible to obtain exact scattering amplitudes of integrable models with $O(N)$ symmetry in $2d$ directly from the underlying algebraic formalism \cite{zamzam}. One of the models for which the so-called Zamolodchikov-Faddeev algebraic formalsim provides the exact S-matrix is the $O(N)$ nonlinear sigma model. Factorization of the S-matrix in this model is demonstrated using the properties of the $1\slash N$ expansion \cite{23} in \cite{25}. Infinite set of conservation laws in this model leads to a more rigorous proof of factorization \cite{26,27}. 

CCFT duals of integrable S-matrices have been studied to a limited extent up to this moment. In the works \cite{cel_2d, kap_trop} it has been demonstrated that rewriting an integrable S-matrix in conformal basis simplifies to taking its Fourier transform. There has been a discussion on possible Fourier space versions of fundamental properties of integrable S-matrices such as factorization, unitarity and crossing symmetry. However, exact expressions for many of the two-dimensional integrable theories are not known. We are aware of the study where a celestial dual for $3d$ $O\left(N\right)$ model has been addressed in the large $N$ limit \cite{cel_ON}.

In this paper we compute non-perturbative celestial amplitudes corresponding to the exact S-matrix of the two-dimensional $O(N)$-symmetric non-linear sigma model. We use the representations of the $O(N)$ model scattering amplitudes as products of Gamma functions and perform their Fourier transform using Mellin transform. The resulting celestial amplitudes are expressed as particular cases of Mejer G-functions, which can be integrated explicitly for all values of symmetry parameter $N$ except for $N=4$. We prove that the Fourier space version of crossing symmetry holds for obtained celestial amplitudes. We also demonstrate that for $N=3$  celestial amplitudes take a particular simplified form.

This paper is organized as follows. In Section 2 we review the fundamental properties of two-dimensional integrable S-matrices and the general approach to computing their CCFT duals. In Section 3 we compute celestial amplitudes corresponding to $O(N)$ sigma model and analyze their properties.

\section{Two-dimensional $\mathbf{O\left(N\right)}$ model. CCFT duals for integrable models}
In this section we describe general properties of scattering matrices with $O(N)$ symmetry. Exact solutions for $O(N)$ nonlinear sigma model with $N\ge 3$ can be obtained from Zamolodchikov-Faddeev algebraic formalism. We also review the unitarity, crossing symmetry and factorization properties of scattering amplitudes. Celestial amplitudes corresponding to integrable S-matrices become their Fourier transform. We briefly review this connection and the general approach to celestial conformal field theory duals for integrable theories. 

\subsection{S-matrix of the O(N) model}

The S-matrices in two-dimensional integrable models are exactly solvable and non-trivial due to the presence of high-order conserved charges \cite{iag}. The $n$-particle S-matrix corresponding to elastic scattering is defined as a unitary linear operator
\begin{equation}
	S_n\left(p_1,\dots,\;p_n\right)={}_{out} {\left\langle p_1,\dots,p_n \;\vert\; p_1,\dots,p_n \right\rangle}_{in}
\end{equation}
The general $n$-particle S-matrix of a two-dimensional integrable model factorizes into the product of $n\left(n-1\right)/2$ two-particle ones \cite{iag}
\begin{equation}
	{{S}_{n}}\left( {{p}_{1}},\ldots,{{p}_{n}} \right)=\prod\limits_{i=1}^{n-1}{\prod\limits_{j=i+1}^{n}{{{S}_{2}}\left( {{p}_{i}},{{p}_{j}} \right)}}.
\end{equation}
In two dimensions, the S-matrix is usually parametrized by the rapidity variable $\theta$:
\begin{equation}
	p_{i}^{0}\,=\,{{m}_{i}}\,\cosh \,{{\theta }_{i}},\qquad p_{i}^{1}\,=\,{{m}_{i}}\,\sinh \,{{\theta }_{i}}.
\end{equation}
Scattering amplitudes are functions of rapidity differences ${{\theta }_{ij}}={{\theta }_{i}}-{{\theta }_{j}}$ due to Lorentz invariance.

Exact expressions for two-dimensional integrable S-matrices can be obtained using an algebraic formalism introduced as follows \cite{zamzam}. Asymptotic states are defined as a set of vertex operators  ${{A}_{{{a}_{i}}}}\left( {{p}_{i}} \right),\;\; {{A}^{\dagger }}^{{{a}_{i}}}\left( {{p}_{i}} \right)$ that generalize creation and annihilation operators
\begin{equation}
	\label{algasymp}
	\begin{aligned}
		& \left| {{\theta }_{1}},{{\theta }_{2}},\ldots {{\theta }_{n}} \right\rangle _{{{a}_{1}},\,{{a}_{2}},\ldots ,{{a}_{n}}}^{in}\,={{A}_{{{a}_{1}}}}\left( {{\theta }_{1}} \right){{A}_{{{a}_{2}}}}\left( {{\theta }_{2}} \right)\ldots {{A}_{{{a}_{n}}}}\left( {{\theta }_{n}} \right)\left| 0 \right\rangle ,\\ 
		& \left| {{\theta }_{1}},{{\theta }_{2}},\ldots {{\theta }_{n}} \right\rangle _{{{a}_{1}},{{a}_{2}},\ldots ,{{a}_{n}}}^{out}={{A}_{{{a}_{n}}}}\left( {{\theta }_{n}} \right)\ldots {{A}_{{{a}_{2}}}}\left( {{\theta }_{2}} \right){{A}_{{{a}_{1}}}}\left( {{\theta }_{1}} \right)\left| 0 \right\rangle, \\ 
		& {}_{{{a}_{1}},{{a}_{2}},\ldots,{{a}_{n}}}^{in}\left\langle  {{\theta }_{1}},{{\theta }_{2}},\ldots {{\theta }_{n}} \right|=\left\langle  0 \right|{{A}^{\dagger }}^{{{a}_{1}}}\left( {{\theta }_{1}} \right)\,{{A}^{\dagger }}^{{{a}_{2}}}\left( {{\theta }_{2}} \right)\ldots {{A}^{\dagger }}^{{{a}_{n}}}\left( {{\theta }_{n}} \right),\\ 
		& {}_{{{a}_{1}},{{a}_{2}},\ldots ,{{a}_{n}}}^{out}\left\langle  {{\theta }_{1}},{{\theta }_{2}},\ldots {{\theta }_{n}} \right|=\left\langle  0 \right|{{A}^{\dagger }}^{{{a}_{n}}}\left( {{\theta }_{n}} \right)\ldots{{A}^{\dagger }}^{{{a}_{2}}}\left( {{\theta }_{2}} \right){{A}^{\dagger }}^{{{a}_{1}}}\left( {{\theta }_{1}} \right). 
	\end{aligned}
\end{equation}
Here the indices $a_i$ label the particle type and decreasing order of rapidities is conventional $\theta_{1} > \theta_{2} > \theta_{3} > \dots > \theta_n$.

The operators ${{A}_{{{a}_{i}}}}\left( {{p}_{i}} \right),\;\; {{A}^{\dagger }}^{{{a}_{i}}}\left( {{p}_{i}} \right)$ form an associative non-commutative algebra called Zamolodchikov-Faddeev algebra \cite{zamzam}.  The definition of asymptotic states given above makes it possible to interpret scattering processes as permutations of corresponding vertex operators. The defining commutation relations of the Zamolodchikov-Faddeev algebra are of the form
\begin{equation}
	\label{algcommrel}
	\begin{aligned}
		{{A}_{i}}\left( {{\theta }_{1}} \right){{A}_{j}}\left( {{\theta }_{2}} \right) &={{A}_{l}}\left( {{\theta }_{2}} \right){{A}_{k}}\left( {{\theta }_{1}} \right)S_{ij}^{kl}\left( {{\theta }_{1}}-{{\theta }_{2}} \right), \\ 
		{{A}^{\dagger i}}\left( {{\theta }_{1}} \right){{A}^{\dagger j }}\left( {{\theta }_{2}} \right) &=S_{kl}^{ij}\left( {{\theta }_{1}}-{{\theta }_{2}} \right){{A}^{\dagger l}}\left( {{\theta }_{2}} \right){{A}^{\dagger k}}\left( {{\theta }_{1}} \right), \\ 
		{{A}^{\dagger k}}\left( {{\theta }_{1}} \right){{A}_{j}}\left( {{\theta }_{2}} \right) &={{A}_{l}}\left( {{\theta }_{2}} \right)S_{ij}^{kl}\left( {{\theta }_{2}}-{{\theta }_{1}} \right){{A}^{\dagger i}}\left( {{\theta }_{1}} \right)+\delta \left( {{\theta }_{1}}-{{\theta }_{2}} \right)\delta _{j}^{k}. 
	\end{aligned}
\end{equation}
These relations generalize the usual bosonic and fermionic commutation relations.

The spectrum of an integrable model possessing $O(N)$ isotopic symmetry is given by $n$ partices ${{A}_{i}},\quad i=1,2,\ldots N$ of equal masses $m$ \cite{zamzam}.

The general S-matrix takes the form
\begin{equation}
	\begin{aligned}
		\label{s_on}
		_{ik}{{S}_{jl}}&=\left\langle {{A}_{j}}\left( {{p}_{1}}^{\prime } \right){{A}_{l}}\left( {{p}_{2}}^{\prime } \right),out|{{A}_{i}}\left( {{p}_{1}} \right){{A}_{k}}\left( {{p}_{2}} \right),in \right\rangle\\ 
		&=\delta \left( {{p}_{1}}-{{p}_{1}}^{\prime } \right)\delta \left( {{p}_{2}}-{{p}_{2}}^{\prime } \right)\left[ {{\delta }_{ik}}{{\delta }_{jl}}{{S}_{1}}\left( s \right)+{{\delta }_{ij}}{{\delta }_{kl}}{{S}_{2}}\left( s \right)+{{\delta }_{il}}{{\delta }_{jk}}{{S}_{3}}\left( s \right) \right] \\
		&\hspace{10pt}\pm \delta \left( {{p}_{2}}-{{p}_{1}}^{\prime } \right)\delta \left( {{p}_{1}}-{{p}_{2}}^{\prime } \right)\left[ {{\delta }_{ik}}{{\delta }_{jl}}{{S}_{1}}\left( s \right)+{{\delta }_{kj}}{{\delta }_{il}}{{S}_{2}}\left( s \right)+{{\delta }_{kl}}{{\delta }_{ij}}{{S}_{3}}\left( s \right) \right],
	\end{aligned}
\end{equation}
where $s$ is the usual Mandelstam variable $s=\left(p_1+p_2\right)^2$ and $s=4{{m}^{2}}\,{{\cosh }^{2}}\left( {\theta }/{2}\right)$.
Corresponding commutation relations for the operators ${{A}_{i}}\left( \theta  \right),\quad i=1,2,\ldots, N$ are
\begin{multline}
	\label{comm}
	{{A}_{i}}\left( {{\theta }_{1}} \right){{A}_{j}}\left( {{\theta }_{2}} \right)={{\delta }_{ij}}{{S}_{1}}\left( {{\theta }_{12}} \right)\sum\nolimits_{k=1}^{N}{{{A}_{k}}\left( {{\theta }_{2}} \right){{A}_{k}}}\left( {{\theta }_{1}} \right),\\
	+{{S}_{2}}\left( {{\theta }_{12}} \right){{A}_{j}}\left( {{\theta }_{2}} \right){{A}_{i}}\left( {{\theta }_{1}} \right)
	+{{S}_{3}}\left( {{\theta }_{12}} \right){{A}_{i}}\left( {{\theta }_{2}} \right){{A}_{j}}\left( {{\theta }_{1}} \right).
\end{multline}
In this article we restrict our consideration to the $O\left(N\right)$ symmetry with $N\ge3$. Following the conventions of \cite{zamzam}, we denote the amplitudes $S_1,\; S_2$ and $S_3$ by $\sigma_1,\; \sigma_2$ and $\sigma_3$, respectively.

Considering different permutations of operators $A_i \left(\theta\right)$ and using commutation relations given above leads to the following factorization equations for $N\ge 3$
\begin{equation}
	\label{fact}
	\begin{aligned}
		\sigma_3\sigma_2\sigma_3&=\sigma_2\sigma_3\sigma_3+\sigma_3\sigma_3\sigma_2,\\
		\sigma_3\sigma_1\sigma_2&=\sigma_2\sigma_1\sigma_1+\sigma_3\sigma_2\sigma_1,\\
		\sigma_3\sigma_1\sigma_3&=N\sigma_1\sigma_3\sigma_1+\sigma_1\sigma_3\sigma_2+\sigma_1\sigma_3\sigma_3+\sigma_1\sigma_2\sigma_1+\sigma_2\sigma_3\sigma_1+\sigma_3\sigma_3\sigma_1+\sigma_1\sigma_1\sigma_1.
	\end{aligned}
\end{equation}
Here the arguments are given by $\theta ,\; \theta +{\theta }',\; {\theta }',$ respectively in each term.
Another property that can be deduced from permuting operators is unitarity
\begin{multline}
	\label{unit}
	{{\sigma}_{2}}\left( \theta  \right){{\sigma}_{2}}\left( -\theta  \right)+{{\sigma}_{3}}\left( \theta  \right){{\sigma}_{3}}\left( -\theta  \right)=1,\\ 
	\shoveleft{{{\sigma}_{2}}\left( \theta  \right){{\sigma}_{3}}\left( -\theta  \right)+{{\sigma}_{2}}\left( -\theta  \right){{\sigma}_{3}}\left( \theta  \right)=1},\\ 
	\shoveleft{N{{\sigma}_{1}}\left( \theta  \right){{\sigma}_{1}}\left( -\theta  \right)+{{\sigma}_{1}}\left( \theta  \right){{\sigma}_{2}}\left( -\theta  \right)+{{\sigma}_{1}}\left( \theta  \right){{\sigma}_{3}}\left( -\theta  \right)}\\
	+{{\sigma}_{2}}\left( \theta  \right){{\sigma}_{1}}\left( -\theta  \right)+{{\sigma}_{3}}\left( \theta  \right){{\sigma}_{1}}\left( -\theta  \right)=0.
\end{multline}
Crossing relations take the form
\begin{equation}
	\label{cross}
	\begin{aligned}
		& {{\sigma}_{2}}\left( \theta  \right)={{\sigma}_{2}}\left( i\pi -\theta  \right),\\ 
		& {{\sigma}_{1}}\left( \theta  \right)={{\sigma}_{3}}\left( i\pi -\theta  \right).
	\end{aligned}
\end{equation}
It is important to note that crossing relation, unlike unitarity and factorization properties, is not obtained from algebraic formalism but corresponds to common physical considerations.

Fundamental solutions of the factorization equations for two-dimensional $O\left(N\right)$-symmetric intgerable models where $N\ge 3$ are \cite{zamzam}
\begin{equation}
	\label{sol3}
	\begin{aligned}
		{{\sigma }_{3}}\left( \theta  \right)&=-\frac{i\lambda }{\theta }{{\sigma }_{2}}\left( \theta  \right) \\ 
		{{\sigma }_{1}}\left( \theta  \right)&=-\frac{i\lambda }{i\left[ \left( N-2 \right)/2 \right]\lambda-\theta }{{\sigma }_{2}}\left( \theta  \right) \\ 
	\end{aligned}
\end{equation}
The restriction
\begin{equation}
	\label{fromc}
	\lambda =\frac{2\pi }{N-2}
\end{equation}
follows from crossing symmetry.

There are two minimal solutions for ${N\ge 3}$ that coincide when $N=3,4$
\begin{equation}
	\label{minsol3}
	\begin{aligned}
		{{\sigma }_{2}}^{\left( \pm  \right)}\left( \theta  \right)&={{Q}^{\left( \pm  \right)}}\left( \theta  \right){{Q}^{\left( \pm  \right)}}\left( i\pi -\theta  \right),\\
		{{Q}^{\left( \pm  \right)}}\left( \theta  \right)&=\frac{\Gamma \left( \pm \frac{\lambda }{2\pi }-i\frac{\theta }{2\pi } \right)\Gamma \left( \frac{1}{2}-i\frac{\theta }{2\pi } \right)}{\Gamma \left( \frac{1}{2}\pm \frac{\lambda }{2\pi }-i\frac{\theta }{2\pi } \right)\Gamma \left( -i\frac{\theta }{2\pi } \right)}.\\
	\end{aligned}
\end{equation}
Choosing $\sigma_2^{\left(+\right)}$ and omitting the $^{\left(+\right)}$ label, we write the solutions as
\begin{equation}
	\label{sol_plus}
	\begin{aligned}
		{{\sigma }_{1}}\left( \theta  \right)&=-\frac{i\lambda }{i\left[ \left( N-2 \right)/2 \right]\lambda-\theta }{{\sigma }_{2}}\left( \theta  \right), \\ 
		\sigma_2\left(\theta\right)&=\frac{\Gamma \left( \frac{\lambda }{2\pi }-\frac{i\theta }{2\pi } \right)\Gamma \left( \frac{1}{2}-\frac{i\theta }{2\pi } \right)\Gamma \left( \frac{1}{2}+\frac{\lambda }{2\pi }+\frac{i\theta }{2\pi } \right)\Gamma \left( 1+\frac{i\theta }{2\pi } \right)}{\Gamma \left( \frac{1}{2}+\frac{\lambda }{2\pi }-\frac{i\theta }{2\pi } \right)\Gamma \left( -\frac{i\theta }{2\pi } \right)\Gamma \left( 1+\frac{\lambda }{2\pi }+\frac{i\theta }{2\pi } \right)\Gamma \left( \frac{1}{2}+\frac{i\theta }{2\pi } \right)}, \\
		{{\sigma }_{3}}\left( \theta  \right)&=-\frac{i\lambda }{\theta }{{\sigma }_{2}}\left( \theta  \right).
	\end{aligned}
\end{equation}
Expressing all of the amplitudes as products of Gamma functions, we get
\setlength{\jot}{4pt}
\begin{equation}
	\label{sol_plus}
	\begin{aligned}
		{{\sigma }_{1}}\left( \theta  \right)&=-\alpha\frac{\Gamma \left( \alpha-\frac{i\theta }{2\pi } \right)\Gamma \left( \frac{1}{2}-\frac{i\theta }{2\pi } \right)\Gamma \left( \frac{1}{2}+\alpha+\frac{i\theta }{2\pi } \right)\Gamma \left( 1+\frac{i\theta }{2\pi } \right)}{\Gamma \left( \frac{1}{2}+\alpha-\frac{i\theta }{2\pi } \right)\Gamma \left( -\frac{i\theta }{2\pi } \right)\Gamma \left( 1+\alpha+\frac{i\theta }{2\pi } \right)\Gamma\left(\frac{3}{2}\alpha+\frac{i\theta}{2\pi}\right)}, \\ 
		\sigma_2\left(\theta\right)&=\frac{\Gamma \left( \alpha-\frac{i\theta }{2\pi } \right)\Gamma \left( \frac{1}{2}-\frac{i\theta }{2\pi } \right)\Gamma \left( \frac{1}{2}+\alpha+\frac{i\theta }{2\pi } \right)\Gamma \left( 1+\frac{i\theta }{2\pi } \right)}{\Gamma \left( \frac{1}{2}+\alpha-\frac{i\theta }{2\pi } \right)\Gamma \left( -\frac{i\theta }{2\pi } \right)\Gamma \left( 1+\alpha+\frac{i\theta }{2\pi } \right)\Gamma \left( \frac{1}{2}+\frac{i\theta }{2\pi } \right)}, \\
		{{\sigma }_{3}}\left( \theta  \right)&=-\alpha\frac{\Gamma \left( \alpha-\frac{i\theta }{2\pi } \right)\Gamma \left( \frac{1}{2}-\frac{i\theta }{2\pi } \right)\Gamma \left( \frac{1}{2}+\alpha+\frac{i\theta }{2\pi } \right)\Gamma \left( 1+\frac{i\theta }{2\pi } \right)}{\Gamma \left( \frac{1}{2}+\alpha-\frac{i\theta }{2\pi } \right)\Gamma \left(1 -\frac{i\theta }{2\pi } \right)\Gamma \left( 1+\alpha+\frac{i\theta }{2\pi } \right)\Gamma \left( \frac{1}{2}+\frac{i\theta }{2\pi } \right)},
	\end{aligned}
\end{equation} 
\setlength{\jot}{2pt}
where $\alpha\equiv \lambda\slash 2\pi=1\slash\left(N-2\right)$ is introduced for convenience.

It has been shown in \cite{zamzam} that these solutions correspond to the $O\left(N\right)$ symmetric nonlinear sigma model given by the Lagrangian and constraint
\[\mathcal{L}=\frac{1}{2g_0} \sum^N_{i=1} \left(\partial_\mu n_i \right)^2 ; \quad \sum^N_{i=1}n_i^2=1,\]
where $g_0$ is a coupling constant.

\subsection{Integrable S-matrix in conformal basis}
Here we describe the general construction of celestial amplitudes \cite{conf_bas} and illustrate the simplification for the 2$d$-integrable case.

The massive scalar conformal primary wavefunction $\phi_\Delta\left(X^\mu;\vec{\omega}\right)$ of mass $m$ in $\mathbb{R}^{1,\: d+1}$ is a wavefunction labelled by a conformal dimension $\Delta$ and a point $\vec{\omega}$ in $\mathbb{R}^d$. The Fourier expansion of the conformal primary wavefunction is
\begin{equation}
	\label{prim_exp}
	\phi_{\Delta}^{\pm}\left(X^\mu;\vec{\omega}\right)=\int_{H_{d+1}}G_{\Delta}\left(\hat{p};\vec{\omega}\right)\exp\left[\pm im\hat{p}\cdot X\right]\left[d\hat{p}\right],
\end{equation}
where the on-shell momenta, a unit timelike vector $\hat{p}\left(y,\vec{z}\right)$ satisfying $\hat{p}^2=-1$
\begin{equation} \hat{p}^\mu_i\equiv\hat{p}^\mu\left(\theta_i\right)=\frac{p_i}{m}=\left(\cosh\theta_i,\;\sinh\theta_i\right)
\end{equation}
can be parametrized using the coordinates $y,\;\vec{z}$ on $H_{d+1}$ with $y>0$ and $\vec{z}\in \mathbb{R}^d$ as
\begin{equation}
	\label{moment_par}	
	\hat{p}\left(y,\vec{z}\right)=\left(\frac{1+y^2+\vert\vec{z}\vert^2}{2y}, \frac{\vec{z}}{y},\frac{1-y^2-\vert\vec{z}\vert^2}{2y}\right),
\end{equation}
and $\left[d\hat{p}\right]$ is the $SO\left(1,d+1\right)$ invariant measure on $H_{d+1}$
\begin{equation}
	\label{meas}
	\int_{H_{d+1}}\left[d\hat{p}\right]\equiv\int_{0}^\infty \frac{dy}{y^{d+1}}\int d^d\vec{z}=\int\frac{d^{d+1}\hat{p}^i}{\hat{p}^0},\quad i=1,2,\dots,d+1,\quad \hat{p}^0=\sqrt{\hat{p}^i \hat{p}^i +1}.
\end{equation}
The scalar bulk-to-boundary propagator $G_{\Delta}\left(\hat{p};\vec{\omega}\right)$ in $H_{d+1}$ is given by
\begin{equation}
	\label{prop_1}
	G_{\Delta}\left(\hat{p};\vec{\omega}\right)=\left(\frac{y}{y^2+|\vec{z}-\vec{\omega}|^2}\right)^{\Delta},
\end{equation}
where $\vec{\omega}\in\mathbb{R}^d$ lies on the boundary of $H_{d+1}$. Given a map from $\mathbb{R}^d$ to a unit null momentum $q^\mu$ in $\mathbb{R}^{1,\: d+1}$
\begin{equation}
	\label{null_moment}
	q^{\mu}\left(\vec{\omega}\right)=\left(1+|\vec{\omega}|^2,\; 2\vec{\omega}, \; 1-|\vec{\omega}|^2\right),
\end{equation}
the scalar bulk-to-boundary propagator can be parametrized in terms of $\hat{p}^\mu\left(y,\vec{z}\right)$ and $q^\mu\left(\vec{\omega}\right)$ as
\begin{equation}
	\label{prop_2}
	G_{\Delta}\left(\hat{p};q\right)=\frac{1}{\left(-\hat{p}\cdot q\right)^\Delta}.
\end{equation}
The change of basis (\ref{prim_exp}) can be extended to scattering amplitudes in momentum space $\mathcal{A}\left(p_i^{\mu}\right)$ as an integral transform to the basis of conformal primary wavefunctions
\begin{equation}
	\label{amplit_conf_bas}
	\tilde{\mathcal{A}}\left(\delta_i,\vec{\omega}_i\right)\equiv\prod_{k=1}^{n}\int_{H_{d+1}} G_{\Delta_k}\left(\hat{p}_k ; \vec{\omega}_k\right) \mathcal{A}\left(\pm m_i \hat{p}_i^\mu\right)\left[d\hat{p}_k\right],
\end{equation}
and the quantity $\tilde{\mathcal{A}}\left(\delta_i,\vec{\omega}_i\right)$ is known as the massive celestial amplitude. Due to the conformal invariance of the primary wavefunctions, the celestial amplitude transforms covariantly as a $d$-dimensional CFT $n$-point function.

It has been shown in \cite{cel_2d, kap_trop} that the celestial 4-point amplitude for $2\to2$ scattering of massive scalar particles in $2d$ 
\begin{equation}
	\label{2d_cel}
	\mathcal{A}=\left(\prod_{i=1}^4 \int\frac{d\hat{p}^1_i}{\hat{p}^0_i}\right)\prod^4_{i=1}G_{\Delta_i}\left(\hat{p}_i\right) S_{2\to2},
\end{equation}
where $G_{\Delta_i}\left(\hat{p}_i\right)$ is the bulk-to-boundary propagator in $H_1$,
simplifies to a Fourier transform of the S-matrix with respect to rapidity
\begin{equation}
	\label{fourier_ampl}
	\mathcal{A}\left(\omega\right)=\int_{-\infty}^{\infty} e^{i\omega\theta} S\left(\theta\right)d\theta.
\end{equation}
Therefore, in order to compute celestial amplitudes corresponding to the nonlinear sigma model, we need to choose a suitable technique for performing Fourier transform.

\section{Celestial amplitudes dual to $\mathbf{O\left(N\right)}$ model S-matrix}
In this section we compute celestial amplitudes dual to scattering amplitudes of the $O(N)$ nonlinear sigma model. We use the representation of scattering amplitudes as products of Gamma functions. Celestial amplitudes are their Fourier transforms that we obtain using Mellin transform. The resulting celestial amplitudes are expressed as Mejer G-functions. We prove crossing symmetry in Fourier space for obtained celestial amplitudes. We also review simplifications arising in $O(3)$ symmertric case.
\subsection{Celestial amplitudes as Mejer G-functions}
In order to perform the Fourier transform of scattering amplitudes expressed as products of Gamma functions, we recall the following connection between Fourier and Mellin transforms \cite{titch}
\begin{equation}
	\label{mellin_fourier}
	\mathcal{F}\left(it\right)=\int_{-\infty}^{\infty}f\left(e^\xi\right)e^{it\xi}d\xi,
\end{equation}
where the Mellin transform of a function $f\left(x\right)$ is defined as 
\begin{equation}
	\label{mellin_def}
	\mathcal{F}\left(s\right)=\int_0^x f\left(x\right) x^{s-1}dx,
\end{equation}
and $f\left(x\right)$ is given by the inverse Mellin transform, respectively
\begin{equation}
	\label{inv_mellin_def}
	f\left(x\right)=\frac{1}{2\pi i}\int_{c-\infty}^{c+\infty}\mathcal{F}\left(s\right)x^{-s}ds.
\end{equation}
Using the definition of the celestial amplitude corresponding to $2d$-integrable model (\ref{fourier_ampl}), as well as the expression for $\sigma_2$ (\ref{sol_plus}), we get
\begin{equation}
	\label{a2_plus_gen}
	A_2\left(-\frac{\ln x}{2\pi}\right)=\frac{1}{2\pi i}\int\limits_C \frac{\Gamma \left( \alpha+\theta \right)\Gamma \left( \frac{1}{2}+\theta\right)\Gamma \left( \frac{1}{2}+\alpha-\theta \right)\Gamma \left( 1-\theta \right)}{\Gamma \left( \frac{1}{2}+\alpha+\theta \right)\Gamma \left(\theta \right)\Gamma \left( 1+\alpha-\theta\right)\Gamma \left(\frac{1}{2}-\theta\right)} x^\theta d\theta,
\end{equation}
where we have denoted the celestial amplitude corresponding to the $\sigma_2\left(\theta\right)$ solution by $A_2\left(-\frac{\ln x}{2\pi}\right)$.
Similarly, for $A_1\left(-\frac{\ln x}{2\pi}\right)$ we have
\begin{equation}
	\label{a1_plus_gen}
	A_1\left(-\frac{\ln x}{2\pi}\right)=\frac{-\alpha}{2\pi i}\int\limits_C \frac{\Gamma \left( \alpha+\theta \right)\Gamma \left( \frac{1}{2}+\theta\right)\Gamma \left( \frac{1}{2}+\alpha-\theta \right)\Gamma \left( 1-\theta \right)}{\Gamma \left( \frac{1}{2}+\alpha+\theta \right)\Gamma \left(\theta \right)\Gamma \left( 1+\alpha-\theta\right)\Gamma \left(\frac{3}{2}-\theta\right)} x^\theta d\theta,
\end{equation}
The amplitude $A_3\left(-\frac{\ln x}{2\pi}\right)$ corresponding to $\sigma_3\left(\theta\right)$ takes the form
\begin{equation}
	\label{a3_plus_gen}
	A_3\left(-\frac{\ln x}{2\pi}\right)=\frac{-\alpha}{2\pi i}\int\limits_C \frac{\Gamma \left( \alpha+\theta \right)\Gamma \left( \frac{1}{2}+\theta\right)\Gamma \left( \frac{1}{2}+\alpha-\theta \right)\Gamma \left( 1-\theta \right)}{\Gamma \left( \frac{1}{2}+\alpha+\theta \right)\Gamma \left(1+\theta \right)\Gamma \left( 1+\alpha-\theta\right)\Gamma \left(\frac{1}{2}-\theta\right)} x^\theta d\theta.
\end{equation}
It turns out that the resulting expressions for the amplitudes can be simplified using Mejer G-function properties described in \cite{bateman}.

The Mejer G-function is given by \cite{bateman}
\begin{equation}
	\label{mejer_def}
	G^{\, m,\, n}_{\, p,\, q}\biggl(\genfrac{}{}{0pt}{}{a_1,\:\dots,\: a_p}{b_1,\:\dots,\: b_p} \bigg\vert x \biggr)=\frac{1}{2\pi i}\int\limits_C \frac{\prod_{j=1}^{m}\Gamma\left(b_j-s\right)\prod_{j=1}^{n}\Gamma\left(1-a_j+s\right)}{\prod_{j=1}^{m+1}\Gamma\left(1-b_j+s\right)\prod_{j=n+1}^{p}\Gamma\left(a_j-s\right)}x^s ds. 
\end{equation}
Therefore, the celestial amplitude 
$A_2\left(-\frac{\ln x}{2\pi}\right)$	(\ref{a2_plus_gen}) is expressed as Mejer G-function with parameters
\begin{equation}
	\label{a2_plus_mejer}
	A_2\left(-\frac{\ln x}{2\pi}\right)=G^{\, 2,\, 2}_{\,4,\, 4}\biggl(\genfrac{}{}{0pt}{}{a_1,\: a_2,\: a_3,\: a_4}{b_1,\: b_2,\: b_3,\: b_4} \bigg\vert x \biggr),
\end{equation}
where
\[\begin{aligned}
	&b_1=\frac{1}{2}+\alpha, \quad b_2=1, \quad b_3=\frac{1}{2}-\alpha,\quad  b_4=1;\\
	&a_1=1-\alpha, \quad a_2=\frac{1}{2}, \quad a_3=1+\alpha,\quad a_4=\frac{1}{2}.
	\end{aligned}\]
Similarly, the celestial amplitude 
$A_1\left(-\frac{\ln x}{2\pi}\right)$	(\ref{a1_plus_gen}) is expressed via Mejer G-function with parameters
\begin{equation}
	\label{a1_plus_mejer}
	A_1\left(-\frac{\ln x}{2\pi}\right)=-\alpha \; G^{\,2,\, 2}_{\,4,\, 4}\biggl(\genfrac{}{}{0pt}{}{a_1,\: a_2,\: a_3,\: a_4}{b_1,\: b_2,\: b_3,\: b_4} \bigg\vert x \biggr),
\end{equation}
where
\[\begin{aligned}
	&b_1=\frac{1}{2}+\alpha, \quad b_2=1, \quad b_3=\frac{1}{2}-\alpha,\quad  b_4=1;\\
	&a_1=1-\alpha, \quad a_2=\frac{1}{2}, \quad a_3=1+\alpha,\quad a_4=\frac{3}{2}.
\end{aligned}\]
Similarly, the celestial amplitude 
$A_3\left(-\frac{\ln x}{2\pi}\right)$	(\ref{a3_plus_gen}) is 
\begin{equation}
	\label{a3_plus_mejer}
	A_3\left(-\frac{\ln x}{2\pi}\right)=-\alpha \; G^{\,2,\, 2}_{\,4,\, 4}\biggl(\genfrac{}{}{0pt}{}{a_1,\: a_2,\: a_3,\: a_4}{b_1,\: b_2,\: b_3,\: b_4} \bigg\vert x \biggr),
\end{equation}
where
\[\begin{aligned}
	&b_1=\frac{1}{2}+\alpha, \quad b_2=1, \quad b_3=\frac{1}{2}-\alpha,\quad  b_4=0;\\
	&a_1=1-\alpha, \quad a_2=\frac{1}{2}, \quad a_3=1+\alpha,\quad a_4=\frac{1}{2}.
\end{aligned}\]
The integration in this case can be performed using \cite{bateman} \setlength{\jot}{6pt}
\begin{multline}
	\label{G2244_int_l1}
G^{\,2,\, 2}_{\,4,\, 4}\biggl(\genfrac{}{}{0pt}{}{a_1,\: a_2,\: a_3,\: a_4}{b_1,\: b_2,\: b_3,\: b_4} \bigg\vert x \biggr) \\
	= \frac{\Gamma\left(b_2-b_1\right)\Gamma\left(1+b_1-a_1\right)\Gamma\left(1+b_1-a_2\right)}{\Gamma\left(1+b_1-b_3\right)\Gamma\left(1+b_1-b_4\right)\Gamma\left(a_3-b_1\right)\Gamma\left(a_4-b_1\right)} x^{b_1}\\
	\times \;{}_4 F_3 \biggl( \genfrac{}{}{0pt}{}{1+b_1-a_1,\;\dots,\;1+b_1-a_4}{1+b_1-b_2,\;\dots,\;1+b_1-b_4}\bigg\vert x \biggr)\\
	+ \frac{\Gamma\left(b_1-b_2\right)\Gamma\left(1+b_2-a_1\right)\Gamma\left(1+b_2-a_2\right)}{\Gamma\left(1+b_2-b_3\right)\Gamma\left(1+b_2-b_4\right)\Gamma\left(a_3-b_2\right)\Gamma\left(a_4-b_2\right)} x^{b_2}\\
		\times \;{}_4 F_3 \biggl( \genfrac{}{}{0pt}{}{1+b_2-a_1,\;\dots,\;1+b_2-a_4}{1+b_2-b_1,\;\dots,\;1+b_2-b_4} \bigg\vert x \biggl),\\
	 \left|x\right|<1.
\end{multline} 
\begin{multline} 
	\label{G2244_int_g1}
	G^{\,2,\, 2}_{\,4,\, 4}\biggl(\genfrac{}{}{0pt}{}{a_1,\: a_2,\: a_3,\: a_4}{b_1,\: b_2,\: b_3,\: b_4} \bigg\vert x \biggr)\\
	= \frac{\Gamma\left(a_1-a_2\right)\Gamma\left(1+b_1-a_1\right)\Gamma\left(1+b_2-a_1\right)}{\Gamma\left(1+a_3-a_1\right)\Gamma\left(1+a_4-a_1\right)\Gamma\left(a_1-b_3\right)\Gamma\left(a_1-b_4\right)} x^{a_1-1}\\
	\times \;{}_4 F_3 \biggl( \genfrac{}{}{0pt}{}{1+b_1-a_1,\;\dots,\;1+b_4-a_1}{1+a_2-a_1,\;\dots,\;1+a_4-a_1} \bigg\vert x^{-1} \biggr)\\
	+ \frac{\Gamma\left(a_2-a_1\right)\Gamma\left(1+b_1-a_2\right)\Gamma\left(1+b_2-a_2\right)}{\Gamma\left(1+a_3-a_2\right)\Gamma\left(1+a_4-a_2\right)\Gamma\left(a_2-b_3\right)\Gamma\left(a_2-b_4\right)} x^{a_2-1}\\
	\times \;{}_4 F_3 \biggl( \genfrac{}{}{0pt}{}{1+b_1-a_2,\;\dots,\;1+b_4-a_2}{1+a_1-a_2,\;\dots,\;1+a_4-a_2} \bigg\vert x^{-1} \biggr),\\
	\left|x\right|>1.
\end{multline} \setlength{\jot}{2pt}
Here $_4F_3$ is the generalized hypergeometric series \cite{bateman}.
For $A_2\left(-\frac{\ln x}{2\pi}\right)$	(\ref{a2_plus_gen}) we get \setlength{\jot}{6pt}
\begin{multline}
	\label{a2_plus_int_l1}
	A_2\left(-\frac{\ln x}{2\pi}\right)	= \frac{\Gamma\left({1}\slash{2}-\alpha\right)\Gamma\left({1}\slash{2}+2\alpha\right)\Gamma\left(1+\alpha\right)}{\Gamma\left(1+2\alpha\right)\Gamma\left(1\slash2+\alpha\right)\Gamma\left({1}\slash{2}\right)\Gamma\left(-\alpha\right)} x^{1\slash2+\alpha}\\
	\times\Ffourthree{1\slash2+2\alpha}{1+\alpha}{1\slash2}{1+\alpha}{1\slash2+\alpha}{1+2\alpha}{1\slash2+\alpha}{x} \\
	+ \frac{\Gamma\left(-1\slash2+\alpha\right)\Gamma\left(1+\alpha\right)\Gamma\left(3\slash2\right)}{\Gamma\left(3\slash2+\alpha\right)\Gamma\left(1\right)\Gamma\left(\alpha\right)\Gamma\left(-1\slash2\right)} x\\
		\times \Ffourthree{1+\alpha}{3\slash2}{1-\alpha}{3\slash2}{3\slash2-\alpha}{3\slash2+\alpha}{1}{x}, \\
		\left|x\right|<1.
\end{multline}
\begin{multline}
	\label{a2_plus_int_g1}
	A_2\left(-\frac{\ln x}{2\pi}\right)	= \frac{\Gamma\left(1\slash2-\alpha\right)\Gamma\left(1\slash2+2\alpha\right)\Gamma\left(1+\alpha\right)}{\Gamma\left(1+2\alpha\right)\Gamma\left(1\slash2+\alpha\right)\Gamma\left(1\slash2\right)\Gamma\left(-\alpha\right)} x^{-\alpha}\\
	\times \Ffourthree{1\slash2+2\alpha}{1+\alpha}{1\slash2}{1+\alpha}{1\slash2+\alpha}{1+2\alpha}{1\slash2+\alpha}{x^{-1}}\\
	+ \frac{\Gamma\left(-1\slash2+\alpha\right)\Gamma\left(1+\alpha\right)\Gamma\left(3\slash2\right)}{\Gamma\left(3\slash2+\alpha\right)\Gamma\left(1\right)\Gamma\left(\alpha\right)\Gamma\left(-1\slash2\right)} x^{-1\slash2}\\ 
	\times\Ffourthree{1+\alpha}{3\slash2}{1-\alpha}{3\slash2}{3\slash2-\alpha}{3\slash2+\alpha}{1}{x^{-1}},\\
	\left|x\right|>1.
\end{multline} 
Similarly, $A_1\left(-\frac{\ln x}{2\pi}\right)$	(\ref{a1_plus_gen}) takes the form 
\begin{multline}
	\label{a1_plus_int_l1}
	A_1\left(-\frac{\ln x}{2\pi}\right)	= -\alpha\frac{\Gamma\left({1}\slash{2}-\alpha\right)\Gamma\left({1}\slash{2}+2\alpha\right)\Gamma\left(1+\alpha\right)}{\Gamma\left(1+2\alpha\right)\Gamma\left(1\slash2+\alpha\right)\Gamma\left({1}\slash{2}\right)\Gamma\left(1-\alpha\right)} x^{1\slash2+\alpha}\\
		\times\Ffourthree{1\slash2+2\alpha}{1+\alpha}{1\slash2}{\alpha}{1\slash2+\alpha}{1+2\alpha}{1\slash2+\alpha}{x}\\
	-\alpha \frac{\Gamma\left(-1\slash2+\alpha\right)\Gamma\left(1+\alpha\right)\Gamma\left(3\slash2\right)}{\Gamma\left(3\slash2+\alpha\right)\Gamma\left(1\right)\Gamma\left(\alpha\right)\Gamma\left(1\slash2\right)} x\\
	\times\Ffourthree{1+\alpha}{3\slash2}{1-\alpha}{1\slash2}{3\slash2-\alpha}{3\slash2+\alpha}{1}{x},\\
	\left|x\right|<1.
\end{multline}
\begin{multline}
	\label{a1_plus_int_g1}
	A_1\left(-\frac{\ln x}{2\pi}\right)	=-\alpha \frac{\Gamma\left(1\slash2-\alpha\right)\Gamma\left(1\slash2+2\alpha\right)\Gamma\left(1+\alpha\right)}{\Gamma\left(1+2\alpha\right)\Gamma\left(3\slash2+\alpha\right)\Gamma\left(1\slash2\right)\Gamma\left(-\alpha\right)} x^{-\alpha}\\
	\times\Ffourthree{1\slash2+2\alpha}{1+\alpha}{1\slash2}{1+\alpha}{1\slash2+\alpha}{1+2\alpha}{3\slash2+\alpha}{x^{-1}}\\
	-\alpha \frac{\Gamma\left(-1\slash2+\alpha\right)\Gamma\left(1+\alpha\right)\Gamma\left(3\slash2\right)}{\Gamma\left(3\slash2+\alpha\right)\Gamma\left(2\right)\Gamma\left(\alpha\right)\Gamma\left(-1\slash2\right)} x^{-1\slash2}\\	\times\Ffourthree{1+\alpha}{3\slash2}{1-\alpha}{3\slash2}{3\slash2-\alpha}{3\slash2+\alpha}{2}{x^{-1}},\\
	\left|x\right|>1.
\end{multline} 
The amplitude $A_3\left(-\frac{\ln x}{2\pi}\right)$	(\ref{a3_plus_gen}) is
\begin{multline}
	\label{a3_plus_int_l1}
	A_3\left(-\frac{\ln x}{2\pi}\right)	= -\alpha\frac{\Gamma\left({1}\slash{2}-\alpha\right)\Gamma\left({1}\slash{2}+2\alpha\right)\Gamma\left(1+\alpha\right)}{\Gamma\left(1+2\alpha\right)\Gamma\left(3\slash2+\alpha\right)\Gamma\left({1}\slash{2}\right)\Gamma\left(-\alpha\right)} x^{1\slash2+\alpha}\\
	\times\Ffourthree{1\slash2+2\alpha}{1+\alpha}{1\slash2}{1+\alpha}{1\slash2+\alpha}{1+2\alpha}{3\slash2+\alpha}{x}\\
	-\alpha \frac{\Gamma\left(-1\slash2+\alpha\right)\Gamma\left(1+\alpha\right)\Gamma\left(3\slash2\right)}{\Gamma\left(3\slash2+\alpha\right)\Gamma\left(2\right)\Gamma\left(\alpha\right)\Gamma\left(-1\slash2\right)} x\\
	\times\Ffourthree{1+\alpha}{3\slash2}{1-\alpha}{3\slash2}{3\slash2-\alpha}{3\slash2+\alpha}{2}{x},\\
	\left|x\right|<1.
\end{multline}
\begin{multline}
	\label{a3_plus_int_g1}
	A_3\left(-\frac{\ln x}{2\pi}\right)	=-\alpha \frac{\Gamma\left(1\slash2-\alpha\right)\Gamma\left(1\slash2+2\alpha\right)\Gamma\left(1+\alpha\right)}{\Gamma\left(1+2\alpha\right)\Gamma\left(1\slash2+\alpha\right)\Gamma\left(1\slash2\right)\Gamma\left(1-\alpha\right)} x^{-\alpha}\\
	\times\Ffourthree{1\slash2+2\alpha}{1+\alpha}{1\slash2}{\alpha}{1\slash2+\alpha}{1+2\alpha}{1\slash2+\alpha}{x^{-1}}\\
	-\alpha \frac{\Gamma\left(-1\slash2+\alpha\right)\Gamma\left(1+\alpha\right)\Gamma\left(3\slash2\right)}{\Gamma\left(3\slash2+\alpha\right)\Gamma\left(1\right)\Gamma\left(\alpha\right)\Gamma\left(1\slash2\right)} x^{-1\slash2}\\	\times\Ffourthree{1+\alpha}{3\slash2}{1-\alpha}{1\slash2}{3\slash2-\alpha}{3\slash2+\alpha}{1}{x^{-1}},\\
	\left|x\right|>1.
\end{multline}
Changing the variable
\[-\frac{\ln x}{2\pi}=\omega,\]
we get for $A_2\left(\omega\right)$ 
\begin{multline}
	\label{a2_plus_int}
	A_2\left(\omega\right)	=\frac{\Gamma\left({1}\slash{2}-\alpha\right)\Gamma\left({1}\slash{2}+2\alpha\right)\Gamma\left(1+\alpha\right)}{\Gamma\left(1+2\alpha\right)\Gamma\left(1\slash2+\alpha\right)\Gamma\left({1}\slash{2}\right)\Gamma\left(-\alpha\right)} e^{-\pi \omega-2\pi\omega\alpha}H\left(\omega\right) \\
	\times\Ffourthree{1\slash2+2\alpha}{1+\alpha}{1\slash2}{1+\alpha}{1\slash2+\alpha}{1+2\alpha}{1\slash2+\alpha}{e^{-2\pi\omega}}\\
	+\frac{\Gamma\left(-1\slash2+\alpha\right)\Gamma\left(1+\alpha\right)\Gamma\left(3\slash2\right)}{\Gamma\left(3\slash2+\alpha\right)\Gamma\left(1\right)\Gamma\left(\alpha\right)\Gamma\left(-1\slash2\right)} e^{-2\pi\omega}H\left(\omega\right)\\
	\times\Ffourthree{1+\alpha}{3\slash2}{1-\alpha}{3\slash2}{3\slash2-\alpha}{3\slash2+\alpha}{1}{e^{-2\pi\omega}}\\
	+\frac{\Gamma\left(1\slash2-\alpha\right)\Gamma\left(1\slash2+2\alpha\right)\Gamma\left(1+\alpha\right)}{\Gamma\left(1+2\alpha\right)\Gamma\left(1\slash2+\alpha\right)\Gamma\left(1\slash2\right)\Gamma\left(-\alpha\right)} e^{2\pi\omega\alpha}H\left(-\omega\right)\\
		\times \Ffourthree{1\slash2+2\alpha}{1+\alpha}{1\slash2}{1+\alpha}{1\slash2+\alpha}{1+2\alpha}{1\slash2+\alpha}{e^{2\pi\omega}}\\
	+ \frac{\Gamma\left(-1\slash2+\alpha\right)\Gamma\left(1+\alpha\right)\Gamma\left(3\slash2\right)}{\Gamma\left(3\slash2+\alpha\right)\Gamma\left(1\right)\Gamma\left(\alpha\right)\Gamma\left(-1\slash2\right)} e^{\pi\omega}H\left(-\omega\right)\\
	\times\Ffourthree{1+\alpha}{3\slash2}{1-\alpha}{3\slash2}{3\slash2-\alpha}{3\slash2+\alpha}{1}{e^{2\pi\omega}},
\end{multline} 
where $H\left(\omega\right)$ is the Heaviside step function.
The amplitude $A_1\left(\omega\right)$ takes the form 
\begin{multline}
	\label{a1_plus_int}
	A_1\left(\omega\right)	= -\alpha\frac{\Gamma\left({1}\slash{2}-\alpha\right)\Gamma\left({1}\slash{2}+2\alpha\right)\Gamma\left(1+\alpha\right)}{\Gamma\left(1+2\alpha\right)\Gamma\left(1\slash2+\alpha\right)\Gamma\left({1}\slash{2}\right)\Gamma\left(1-\alpha\right)} e^{-\pi\omega-2\pi\omega\alpha}H\left(\omega\right)\\
\Ffourthree{1\slash2+2\alpha}{1+\alpha}{1\slash2}{\alpha}{1\slash2+\alpha}{1+2\alpha}{1\slash2+\alpha}{e^{-2\pi\omega}}\\
	-\alpha \frac{\Gamma\left(-1\slash2+\alpha\right)\Gamma\left(1+\alpha\right)\Gamma\left(3\slash2\right)}{\Gamma\left(3\slash2+\alpha\right)\Gamma\left(1\right)\Gamma\left(\alpha\right)\Gamma\left(1\slash2\right)} e^{-2\pi\omega}H\left(\omega\right)\\
\times\Ffourthree{1+\alpha}{3\slash2}{1-\alpha}{1\slash2}{3\slash2-\alpha}{3\slash2+\alpha}{1}{e^{-2\pi\omega}},\\
	-\alpha \frac{\Gamma\left(1\slash2-\alpha\right)\Gamma\left(1\slash2+2\alpha\right)\Gamma\left(1+\alpha\right)}{\Gamma\left(1+2\alpha\right)\Gamma\left(3\slash2+\alpha\right)\Gamma\left(1\slash2\right)\Gamma\left(-\alpha\right)} e^{2\pi\omega\alpha}H\left(-\omega\right)\\
	\times\Ffourthree{1\slash2+2\alpha}{1+\alpha}{1\slash2}{1+\alpha}{1\slash2+\alpha}{1+2\alpha}{3\slash2+\alpha}{e^{2\pi\omega}}\\
	-\alpha \frac{\Gamma\left(-1\slash2+\alpha\right)\Gamma\left(1+\alpha\right)\Gamma\left(3\slash2\right)}{\Gamma\left(3\slash2+\alpha\right)\Gamma\left(2\right)\Gamma\left(\alpha\right)\Gamma\left(-1\slash2\right)} e^{\pi\omega}H\left(-\omega\right)\\
	\times\Ffourthree{1+\alpha}{3\slash2}{1-\alpha}{3\slash2}{3\slash2-\alpha}{3\slash2+\alpha}{2}{e^{2\pi\omega}}
\end{multline}
and the amplitude $A_3\left(\omega\right)$ is
\begin{multline}
	\label{a3_plus_int}
	A_3\left(\omega\right)	= -\alpha\frac{\Gamma\left({1}\slash{2}-\alpha\right)\Gamma\left({1}\slash{2}+2\alpha\right)\Gamma\left(1+\alpha\right)}{\Gamma\left(1+2\alpha\right)\Gamma\left(3\slash2+\alpha\right)\Gamma\left({1}\slash{2}\right)\Gamma\left(-\alpha\right)} e^{-\pi\omega-2\pi\omega\alpha}H\left(\omega\right)\\
	\times\Ffourthree{1\slash2+2\alpha}{1+\alpha}{1\slash2}{1+\alpha}{1\slash2+\alpha}{1+2\alpha}{3\slash2+\alpha}{e^{-2\pi\omega}}\\
	-\alpha \frac{\Gamma\left(-1\slash2+\alpha\right)\Gamma\left(1+\alpha\right)\Gamma\left(3\slash2\right)}{\Gamma\left(3\slash2+\alpha\right)\Gamma\left(2\right)\Gamma\left(\alpha\right)\Gamma\left(-1\slash2\right)} e^{-2\pi\omega}H\left(\omega\right)\\
	\times\Ffourthree{1+\alpha}{3\slash2}{1-\alpha}{3\slash2}{3\slash2-\alpha}{3\slash2+\alpha}{2}{e^{-2\pi\omega}}\\
	-\alpha \frac{\Gamma\left(1\slash2-\alpha\right)\Gamma\left(1\slash2+2\alpha\right)\Gamma\left(1+\alpha\right)}{\Gamma\left(1+2\alpha\right)\Gamma\left(1\slash2+\alpha\right)\Gamma\left(1\slash2\right)\Gamma\left(1-\alpha\right)} e^{2\pi\omega\alpha}H\left(-\omega\right)\\
	\times\Ffourthree{1\slash2+2\alpha}{1+\alpha}{1\slash2}{\alpha}{1\slash2+\alpha}{1+2\alpha}{1\slash2+\alpha}{e^{2\pi\omega}}\\
	-\alpha \frac{\Gamma\left(-1\slash2+\alpha\right)\Gamma\left(1+\alpha\right)\Gamma\left(3\slash2\right)}{\Gamma\left(3\slash2+\alpha\right)\Gamma\left(1\right)\Gamma\left(\alpha\right)\Gamma\left(1\slash2\right)} e^{\pi\omega}H\left(-\omega\right)\\
	\times\Ffourthree{1+\alpha}{3\slash2}{1-\alpha}{1\slash2}{3\slash2-\alpha}{3\slash2+\alpha}{1}{ e^{2\pi\omega}}.
\end{multline} \setlength{\jot}{2pt}
We note that this integration is valid for all values of symmetry parameters $N$ except for particular case $N=4$. In this case, either parameters $a_1,\; a_2$ or $b_1,\; b_2$ of the Mejer G-function (\ref{mejer_def}) differ by an integer, and the formulas (\ref{G2244_int_l1}, \ref{G2244_int_g1}) are not valid \cite{bateman}. For $N\ge5$ no such complications arise since the values of $\alpha=1\slash\left(N-2\right)$ decrease for increasing $N$.

\subsection{Properties: Crossing symmetry}
Fourier-space version of crossing symmetry relations
\[\begin{aligned}
	\sigma_2\left(\theta\right)&=\sigma_2\left(i\pi-\theta\right),\\
	\sigma_1\left(\theta\right)&=\sigma_3\left(i\pi-\theta\right)
\end{aligned}\]
is
\begin{equation}
	\label{fourier_cross}
	\begin{aligned}
	A_2\left(\omega\right)&=e^{-\pi\omega}A_2\left(-\omega\right),\\
	A_1\left(\omega\right)&=e^{-\pi\omega}A_3\left(-\omega\right).
\end{aligned}\end{equation}
These relations are satisfied for the celestial amplitudes $A_1\left(\omega\right)$ (\ref{a1_plus_int}), $A_2\left(\omega\right)$ (\ref{a2_plus_int})  and $A_3\left(\omega\right)$ (\ref{a3_plus_int}) obtained via Mellin transform. Namely, \setlength{\jot}{6pt}
\begin{multline} 
	e^{-\pi\omega}A_2\left(-\omega\right)	=\frac{\Gamma\left({1}\slash{2}-\alpha\right)\Gamma\left({1}\slash{2}+2\alpha\right)\Gamma\left(1+\alpha\right)}{\Gamma\left(1+2\alpha\right)\Gamma\left(1\slash2+\alpha\right)\Gamma\left({1}\slash{2}\right)\Gamma\left(-\alpha\right)} e^{2\pi\omega\alpha}H\left(-\omega\right) \\
		\times \Ffourthree{1\slash2+2\alpha}{1+\alpha}{1\slash2}{1+\alpha}{1\slash2+\alpha}{1+2\alpha}{1\slash2+\alpha}{e^{2\pi\omega}}\\
	+\frac{\Gamma\left(-1\slash2+\alpha\right)\Gamma\left(1+\alpha\right)\Gamma\left(3\slash2\right)}{\Gamma\left(3\slash2+\alpha\right)\Gamma\left(1\right)\Gamma\left(\alpha\right)\Gamma\left(-1\slash2\right)} e^{\pi\omega}H\left(-\omega\right)\\
	\times\Ffourthree{1+\alpha}{3\slash2}{1-\alpha}{3\slash2}{3\slash2-\alpha}{3\slash2+\alpha}{1}{e^{2\pi\omega}}\\
	+\frac{\Gamma\left(1\slash2-\alpha\right)\Gamma\left(1\slash2+2\alpha\right)\Gamma\left(1+\alpha\right)}{\Gamma\left(1+2\alpha\right)\Gamma\left(1\slash2+\alpha\right)\Gamma\left(1\slash2\right)\Gamma\left(-\alpha\right)} e^{-\pi\omega-2\pi\omega\alpha}H\left(\omega\right)\\
	\times\Ffourthree{1\slash2+2\alpha}{1+\alpha}{1\slash2}{1+\alpha}{1\slash2+\alpha}{1+2\alpha}{1\slash2+\alpha}{e^{-2\pi\omega}}\\
	+ \frac{\Gamma\left(-1\slash2+\alpha\right)\Gamma\left(1+\alpha\right)\Gamma\left(3\slash2\right)}{\Gamma\left(3\slash2+\alpha\right)\Gamma\left(1\right)\Gamma\left(\alpha\right)\Gamma\left(-1\slash2\right)} e^{-2\pi\omega}H\left(\omega\right)\\
	\times\Ffourthree{1+\alpha}{3\slash2}{1-\alpha}{3\slash2}{3\slash2-\alpha}{3\slash2+\alpha}{1}{e^{-2\pi\omega}}\\
	=A_2\left(\omega\right)
\end{multline} 
and
\begin{multline}
	e^{-\pi\omega}A_1\left(-\omega\right)	= -\alpha\frac{\Gamma\left({1}\slash{2}-\alpha\right)\Gamma\left({1}\slash{2}+2\alpha\right)\Gamma\left(1+\alpha\right)}{\Gamma\left(1+2\alpha\right)\Gamma\left(1\slash2+\alpha\right)\Gamma\left({1}\slash{2}\right)\Gamma\left(1-\alpha\right)} e^{2\pi\omega\alpha}H\left(-\omega\right)\\
	\times\Ffourthree{1\slash2+2\alpha}{1+\alpha}{1\slash2}{\alpha}{1\slash2+\alpha}{1+2\alpha}{1\slash2+\alpha}{e^{2\pi\omega}}\\
	-\alpha \frac{\Gamma\left(-1\slash2+\alpha\right)\Gamma\left(1+\alpha\right)\Gamma\left(3\slash2\right)}{\Gamma\left(3\slash2+\alpha\right)\Gamma\left(1\right)\Gamma\left(\alpha\right)\Gamma\left(1\slash2\right)} e^{\pi\omega}H\left(-\omega\right)\\
	\times\Ffourthree{1+\alpha}{3\slash2}{1-\alpha}{1\slash2}{3\slash2-\alpha}{3\slash2+\alpha}{1}{ e^{2\pi\omega}}\\
	-\alpha \frac{\Gamma\left(1\slash2-\alpha\right)\Gamma\left(1\slash2+2\alpha\right)\Gamma\left(1+\alpha\right)}{\Gamma\left(1+2\alpha\right)\Gamma\left(3\slash2+\alpha\right)\Gamma\left(1\slash2\right)\Gamma\left(-\alpha\right)} e^{-\pi\omega-2\pi\omega\alpha}H\left(\omega\right)\\
	\times\Ffourthree{1\slash2+2\alpha}{1+\alpha}{1\slash2}{1+\alpha}{1\slash2+\alpha}{1+2\alpha}{3\slash2+\alpha}{e^{-2\pi\omega}}\\
	-\alpha \frac{\Gamma\left(-1\slash2+\alpha\right)\Gamma\left(1+\alpha\right)\Gamma\left(3\slash2\right)}{\Gamma\left(3\slash2+\alpha\right)\Gamma\left(2\right)\Gamma\left(\alpha\right)\Gamma\left(-1\slash2\right)} e^{-2\pi\omega}H\left(\omega\right)\\
	\times\Ffourthree{1+\alpha}{3\slash2}{1-\alpha}{3\slash2}{3\slash2-\alpha}{3\slash2+\alpha}{2}{e^{-2\pi\omega}}\\
	=A_3\left(\omega\right).
\end{multline} \setlength{\jot}{2pt}
As we have shown, the Fourier space version of crossing symmetry relations holds for the obtained celestial amplitudes. However, we have not succeeded in obtaining analogues of unitarity (\ref{unit}) and factorization (\ref{fact}) properties valid for these results. We argue that this is expected due to the fact that unitarity and factorization in the forms (\ref{unit}) and (\ref{fact}) are obtained directly from the Zamolodchikov-Faddeev algebra with commutation relations (\ref{comm}). In other words, unitarity and factorization properties for the analytical solutions $\sigma_1,\;\sigma_2$ and $\sigma_3$ are given in their algebraical form. Crossing symmetry, however, is formulated commonly without any derivation from the algebraic structure. Therefore, the important underlying question is how to translate the fundamental Zamolodchikov-Faddeev algebraic structure to the Fourier space of celestial amplitudes.

\subsection{Simplifications: O(3) symmetry case}
It turns out there is a particular case where celestial amplitudes have a very simple form. Considering the $O\left(3\right)$ symmetry case, $\alpha=\lambda\slash2\pi=1\slash\left(3-2\right)=1$ and the amplitude $A_2\left(\omega\right)$ takes the form \setlength{\jot}{6pt}
\begin{multline} 
	A_2\left(\omega\right) \\ =\frac{\Gamma\left(-{1}\slash{2}\right)\Gamma\left({5}\slash{2}\right)\Gamma\left(2\right)}{\Gamma\left(3\right)\Gamma\left(3\slash2\right)\Gamma\left({1}\slash{2}\right)\Gamma\left(-1\right)} e^{-3\pi\omega} H\left(\omega\right)\; \Ffourthree{5\slash2}{2}{1\slash2}{2}{3\slash2}{3}{3\slash2}{e^{-2\pi\omega}}\\
	+ \frac{\Gamma\left(1\slash2\right)\Gamma\left(2\right)\Gamma\left(3\slash2\right)}{\Gamma\left(5\slash2\right)\Gamma\left(1\right)\Gamma\left(1\right)\Gamma\left(-1\slash2\right)} e^{-2\pi\omega}H\left(\omega\right)\;\Ffourthree{2}{3\slash2}{0}{3\slash2}{1\slash2}{5\slash2}{1}{e^{-2\pi\omega}}\\
 +\frac{\Gamma\left(-1\slash2\right)\Gamma\left(5\slash2\right)\Gamma\left(2\right)}{\Gamma\left(3\right)\Gamma\left(3\slash2\right)\Gamma\left(1\slash2\right)\Gamma\left(-1\right)} e^{2\pi\omega}H\left(-\omega\right) \;
 \Ffourthree{5\slash2}{2}{1\slash2}{2}{3\slash2}{3}{3\slash2}{e^{2\pi\omega}}\\
	+ \frac{\Gamma\left(1\slash2\right)\Gamma\left(2\right)\Gamma\left(3\slash2\right)}{\Gamma\left(5\slash2\right)\Gamma\left(1\right)\Gamma\left(1\right)\Gamma\left(-1\slash2\right)} e^{\pi\omega}H\left(-\omega\right)\;
	\Ffourthree{2}{3\slash2}{0}{3\slash2}{1\slash2}{5\slash2}{1}{e^{2\pi\omega}}\\
\end{multline} \setlength{\jot}{2pt}
Since the generalized hypergeometric series is defined as \cite{bateman}
\begin{equation}
	\label{gen_hyper_def}
	{}_pF_q\biggl(\genfrac{}{}{0pt}{}{a_1,\:\dots,\: a_p}{\rho_1,\:\dots\:\rho_q} \bigg\vert z \biggr)=\sum_{n=0}^\infty\frac{\left(a_1\right)_n \dots \left(a_p\right)_n}{\left(\rho_1\right)_n \dots \left(\rho_q\right)_n}\frac{z^n}{n!},
\end{equation}
where
\[\left(a\right)_0=1,\quad\left(a\right)_n=a\left(a+1\right) \dots \left(a+n-1\right)=\Gamma\left(a+n\right)\slash\Gamma\left(a\right),\]
it terminates when some $a_k$ is a non-positive integer. Therefore,
\[\Ffourthree{2}{3\slash2}{0}{3\slash2}{1\slash2}{5\slash2}{1}{e^{-2\pi\omega}}=\Ffourthree{2}{3\slash2}{0}{3\slash2}{1\slash2}{5\slash2}{1}{e^{2\pi\omega}}=1.\]
Since $\Gamma\left(-1\right)$ goes to infinity, contributions containing $\Gamma\left(-1\right)$ in denominator cancel in each case and the amplitude is
\begin{multline}
	A_2\left(\omega\right)= \frac{\Gamma\left(1\slash2\right)\Gamma\left(2\right)\Gamma\left(3\slash2\right)}{\Gamma\left(5\slash2\right)\Gamma\left(1\right)\Gamma\left(1\right)\Gamma\left(-1\slash2\right)} e^{-2\pi\omega}H\left(\omega\right)\\
	+\frac{\Gamma\left(1\slash2\right)\Gamma\left(2\right)\Gamma\left(3\slash2\right)}{\Gamma\left(5\slash2\right)\Gamma\left(1\right)\Gamma\left(1\right)\Gamma\left(-1\slash2\right)} e^{\pi\omega}H\left(-\omega\right).
\end{multline}
Due to properties of the Gamma function
\[\Gamma\left(1\right)=\Gamma\left(2\right)=1,\quad \Gamma\left(1\slash2\right)=-1\slash2\,\Gamma\left(-1\slash2\right),\quad \Gamma\left(5\slash2\right)=3\slash 2\,\Gamma\left(3\slash2\right),\]
the amplitude simplifies to
\begin{equation}
	\label{a2_plus_o3}
	A_2\left(\omega\right)=-\frac{1}{3}e^{-2\pi \omega}H\left(\omega\right)-\frac{1}{3}e^{\pi \omega}H\left(-\omega\right).
\end{equation}
Similarly, the amplitude $A_1\left(\omega\right)$ in $O\left(3\right)$-symmetry case is
\begin{equation}
	\label{a1_plus_o3}
	A_1\left(\omega\right)=-\frac{2}{3}e^{-2\pi \omega}H\left(\omega\right)+\frac{1}{3}e^{\pi \omega}H\left(-\omega\right),
\end{equation}
and the amplitude $A_3\left(\omega\right)$ takes the form
\begin{equation}
	\label{a3_plus_o3}
	A_3\left(\omega\right)=\frac{1}{3}e^{-2\pi \omega}H\left(\omega\right)-\frac{2}{3}e^{\pi \omega}H\left(-\omega\right).
\end{equation}
It can easily be seen that these results satisfy the Fourier space crossing symmetry (\ref{fourier_cross}). The simple exponential structure of obtained celestial amplitudes makes it possible to speculate on generalizing relations of the Zamolodchikov-Faddeev algebra to celestial space. However, establishing the general analytical constraints on celestial amplitudes corresponding to $O\left(N\right)$ symmetry is more complicated due to the presence of hypergeometric series with no apparent simplifications of this kind.

\section{Conclusion}
The correspondence between the symmetries characteristic of exact two-dimensional integrable S-matrices and those of their dual celestial CFT correlators has hardly been explored. Establishing a self-consistent set of analytical constraints on celestial amplitudes corresponding to $2d$ integrable theories, preferrably in a form of some algebraic structure, is an open question as well \cite{cel_2d, kap_trop}. Further steps in this direction could shed light on the nature of conformal symmetries present in quantum field theory models \cite{cel_alg, conf_wave_exp}. Another important question is finding the $AdS$ duals of celestial conformal correlators \cite{dual_AdS2, twist_hol}.

In this paper we have computed celestial amplitudes corresponding to one of the two-dimensional integrable models with non-trivial symmetries, the $O(N)$ non-linear sigma model. We have shown that these celestial amplitudes dual to the exact $O\left(N\right)$ S-matrix are particular cases of the Mejer G-function. We have performed the integration for all cases except for $N=4$. In these cases, celestial amplitudes satisfy the Fourier space crossing symmetry condition. We have also obtained simple expressions for conformal correlators for $N=3.$

One of the most interesting questions that remains open is obtaining Fourier space versions of other analytical properties known for integrable scattering amplitudes. We have not been able to show that celestial amplitudes satisfy any of Fourier transformed unitarity or factorization relations written for the $O(N)$ model. Since the form of unitarity and factorization equations that we have considered follows from the Zamolodchikov-Faddeev algebraic setup, it is rather natural not to expect their direct translation to Fourier space. We suggest there is a need to build an analogue of the Zamolodchikov-Faddeev algebraic formalsim in celestial space, that could turn out to be a completely different algebraic structure, in order to obtain a full set of analytical properties for celestial amplitudes.

\section{Acknowledgements}
The author is grateful to I.~E. Shenderovich for enlightening discussions on integrable theories and their conformal duals, to S.~E. Derkachov for pointing out related Mellin-Barnes integration techniques and to A.~G. Pronko for reading the manuscript and useful remarks. This work has been supported in part by the grant 075-15-2022-289 contributed to the Euler International Mathematical Institute.


\begin{thebibliography}{100}
	
	\bibitem{strom_lect}
	 A. Strominger, Lectures on Infrared Structure of Gravity, Princeton University Press (2018);
	 \href{https://arxiv.org/abs/1703.05448}{arXiv:1703.05448}.
	 
	 \bibitem{past_lect}
	 S.~Pasterski, Lectures on celestial amplitudes, The Eur. Phys. J. C, 81 (2021), 1062;
	 \href{https://arXiv.org/abs/2108.04801}{arXiv:2108.04801}.

	 
	 \bibitem{conf_sym}
	 S. Pasterski and S. H. Shao, Conformal basis for flat space amplitudes, Phys. Rev. D 96 (2017), 065022; \href{https://arXiv.org/abs/1705.01027}{arXiv:1705.01027}.
	 
	
	\bibitem{conf_bas}
	S. Pasterski, S. H. Shao, and A. Strominger, Flat space amplitudes and conformal symmetry of the celestial sphere, Phys. Rev. D 96 (2017), 065026; \href{https://arXiv.org/abs/1701.00049}{arXiv:1701.00049}.
	
	 \bibitem{hol}
	C. Cheung, A. de la Fuente and R. Sundrum, 4D scattering amplitudes and asymptotic symmetries from 2D CFT, JHEP 01 (2017), 112; \href{https://arXiv.org/abs/1609.00732}{arXiv:1609.00732}.
	
	\bibitem{boer_solod}
	J. de Boer and S. N. Solodukhin, A Holographic reduction of Minkowski space-time, Nucl. Phys. B665 (2003) 545–593,
	\href{https://arxiv.org/abs/hep-th/0303006}{hep-th/0303006}.
	
	\bibitem{bomb}
	D. Bombardelli, S-matrices and integrability, J. Phys. A: Math. Theor. 49 (2016), 323003; \href{https://arXiv.org/abs/1606.02949}{arXiv:1606.02949}.
	
	
	\bibitem{parke}
	S. Parke, Absence of particle production and factorization of the S-matrix in 1+ 1 dimensional models, Nucl. Phys. B 174 (1980), 166-182.
	
	\bibitem{iag}
	D. Iagolnitzer, Factorization of the multiparticle S matrix in two-dimensional spacetime models, Phys. Rev. D 18 (1976), 1275.	
	
	\bibitem{muss}
	G. Mussardo, Statistical Field Theory, Oxford University Press, Oxford (2009).
	
	\bibitem{zamzam} 
	A. B. Zamolodchikov and A. B. Zamolodchikov, Factorized S-matrices in two dimensions as the exact solutions of certain relativistic quantum field theory models, Annals of Physics 120 (1979), 253-291.
	
	\bibitem{23}
	E.~Br\'ezin and  J. Zinn-Justin, Spontaneous breakdown of continuous symmetries near two dimensions, Phys. Rev. B 14 (1976), 3110.
	
	\bibitem{25}
	A. B. Zamolodchikov and A. B. Zamolodchikov, Relativistic factorized S matrix in two dimensions having O(N) isotopic symmetry, JINR E2-10857 (1977).
	
	\bibitem{26}
	A. M. Polyakov, Hidden symmetry of the two-dimensional chiral fields. Phys. Lett. B, 72 (1977), 224-226.
	
	\bibitem{27}
	I. Y. Arefeva, E. R. Nissimov, S. J. Pacheva, and P. P. Kulish, Infinite set of conservation laws of the quantum chiral field in two-dimensional space-time, LOMI-E-78-1 (1977).
	
	\bibitem{cel_2d}
	S. Duary, Celestial amplitude for 2d theory, JHEP 12 (2022), 1-22; \href{https://arXiv.org/abs/2209.02776}{arXiv:2209.02776}.
	
	\bibitem{kap_trop}
	D. Kapec and A. Tropper, Integrable field theories and their CCFT duals, JHEP 02 (2023), 128;
	\href{https://arXiv.org/abs/2210.16861}{arXiv:2210.16861}.
	
	\bibitem{cel_ON}
	D. Garc\'ia-Sep\'ulveda, A. Guevara, J. Kulp, and J. Wu, Notes on resonances and unitarity from celestial amplitudes, JHEP 09 (2022), 1-40;
	\href{https://arXiv.org/abs/2205.14633}{arXiv:2205.14633}.

	
    \bibitem{titch}
    E. C. Titchmarsh, Introduction to the theory of Fourier integrals, 1948.
    
    \bibitem{bateman}
    H. Bateman, Introduction to higher transcendental functions, Vol. 1, 1953.
    
    \bibitem{cel_alg}
    R. Monteiro, From Moyal deformations to chiral higher-spin theories and to celestial algebras, JHEP 03 (2023), 1-28; 
    \href{https://arxiv.org/abs/2212.11266}{arXiv:2212.11266}.
  
    
    \bibitem{conf_wave_exp}
    C. Liu and D. A. Lowe, Conformal wave expansions for flat space amplitudes, JHEP 07 (2021), 1-18;
    \href{https://arxiv.org/abs/2105.01026}{arXiv:2105.01026}.

    
    \bibitem{dual_AdS2}
    S. Duary, Melting $ AdS_2 $-ice into flatland: flat limit of massless scalar scattering, arXiv preprint
    \href{https://arxiv.org/abs/2305.20037}{arXiv:2305.20037} (2023).
   
    
    \bibitem{twist_hol}
    K. Costello, and N. M. Paquette, Celestial holography meets twisted holography: 4d amplitudes from chiral correlators, JHEP 10 (2022), 1-69;
    \href{https://arxiv.org/abs/2201.02595}{arXiv:2201.02595}.
 


	
\end{thebibliography}
\end{document}